\documentclass[sigconf,nonacm]{acmart}

\usepackage{amsmath}
\usepackage{graphicx}
\usepackage{tikz}
\usepackage{subcaption}
\usepackage{xcolor}
\usepackage{xspace}
\usepackage{url}
\usepackage{float}
\usepackage{multirow}
\usepackage{booktabs}
\usepackage{tabularx}
\usepackage{adjustbox}
\usepackage{longtable}
\usepackage{diagbox}
\usepackage{algorithm}
\usepackage{algpseudocode}
\usepackage{listings}
\usepackage{courier}
\usepackage{colortbl}
\usepackage{acronym}
\usepackage{comment}
\usepackage{array}

\definecolor{dkgreen}{rgb}{0,0.6,0}
\definecolor{gray}{rgb}{0.5,0.5,0.5}
\definecolor{mauve}{rgb}{0.58,0,0.82}

\lstdefinestyle{java}{
  language=Java,
  aboveskip=1mm,
  belowskip=1mm,
  showstringspaces=false,
  columns=flexible,
  basicstyle={\small\ttfamily},
  numbers=left,
  numberstyle=\tiny\color{gray},
  numbersep=5pt,
  keywordstyle=\color{blue},
  commentstyle=\color{dkgreen},
  stringstyle=\color{mauve},
  breaklines=true,
  breakatwhitespace=true,
  tabsize=1
}

\newcommand{\sysname}{\textsf{NeuroTrace}\xspace}

\begin{document}

\title{\sysname{}: Inference Provenance-Based Detection of\\ Adversarial Examples}

\author{Firas Ben Hmida}
\email{fbhmida@umich.edu}
\affiliation{%
  \institution{University of Michigan-Dearborn}
  \city{Dearborn}
  \state{Michigan}
  \country{USA}
}

\author{Philemon Hailemariam}
\email{philemon@umich.edu}
\affiliation{%
  \institution{University of Michigan-Dearborn}
  \city{Dearborn}
  \state{Michigan}
  \country{USA}
}

\author{Kashif Ali Khan}
\email{kashifpk@umich.edu}
\affiliation{%
  \institution{University of Michigan-Dearborn}
  \city{Dearborn}
  \state{Michigan}
  \country{USA}
}

\author{Birhanu Eshete}
\email{birhanu@umich.edu}
\affiliation{%
  \institution{University of Michigan-Dearborn}
  \city{Dearborn}
  \state{Michigan}
  \country{USA}
}



\begin{abstract}

Deep neural networks (DNNs) remain largely opaque at inference time, limiting our ability to detect and diagnose malicious input manipulations such as adversarial examples. Existing detection methods predominantly rely on layer-local signals (e.g., activations or attribution scores), leaving cross-layer information flow and execution structure underexplored. 

We introduce \sysname{}, a framework and open dataset for analyzing \emph{inference provenance} through Inference Provenance Graphs (IPGs). IPGs are heterogeneous graphs that capture both activation behavior and parameter-induced dataflow during a model’s forward pass, providing a structured representation of how information propagates through the network. \sysname{} includes (i) a reproducible extraction engine that instruments model execution, (ii) a standardized graph representation compatible with heterogeneous GNNs, and (iii) a benchmark suite spanning multiple adversarial attack families across vision and malware domains.

Using this framework, we evaluate IPG-based detectors for adversarial example detection under intra-attack, multi-attack, and cross-threat transfer settings. Our results show that inference provenance provides a strong and transferable signal for distinguishing adversarial and benign inputs, achieving consistently high detection performance and improving over prior graph-based baselines. We further analyze the conditions under which provenance-based detection generalizes across attack types, as well as the associated runtime and storage trade-offs. By releasing the dataset, extraction pipeline, and evaluation protocol, \sysname{} enables systematic study of inference-time behavior and establishes inference provenance as a practical foundation for building more transparent and auditable machine learning systems.

\end{abstract}

\maketitle

\section{Introduction}\label{sec:intro}

Deep neural networks (DNNs) underpin critical systems, yet the internal evidence guiding their predictions remains largely opaque. This lack of transparency limits our ability to detect and diagnose failures such as adversarial examples and backdoors. Existing detection approaches typically rely on layer-local signals (e.g., logits, gradients, or saliency maps), which treat model components in isolation and fail to capture how information propagates across layers during inference.

This paper explores a different perspective: \emph{inference provenance}—the structured representation of how input-dependent activations evolve through a model’s computation. Inference Provenance Graphs (IPGs)~\cite{hmida2025deepprovbehavioralcharacterizationrepair} provide such a representation, where nodes encode activation states and edges encode parameter-induced connectivity. Unlike prior approaches that summarize activations, IPGs explicitly model the \emph{information flow} of an inference, capturing both what activates and how signals propagate across the network.

We hypothesize that adversarial perturbations induce \emph{systemic disruptions} in this information flow, beyond localized activation changes. If true, this suggests that inference provenance can serve as a complementary signal for detecting adversarial inputs—one that is less tied to specific attack mechanisms and more reflective of global execution behavior.

However, studying inference provenance in practice remains challenging. Existing work relies on ad hoc extraction pipelines, task-specific graph definitions, and limited experimental setups, making it difficult to evaluate the robustness and generality of provenance-based methods.

To address this gap, we introduce \sysname{}, a unified framework for extracting, representing, and analyzing inference provenance in DNNs, together with an empirical study of its utility for adversarial example detection. \sysname{} provides:
(i) a reproducible extraction pipeline that converts model executions into structured graphs,
(ii) a standardized representation for heterogeneous provenance data, and
(iii) a benchmark suite spanning multiple domains and threat models.

Using this framework, we investigate whether inference provenance constitutes a useful and transferable signal for adversarial detection. Our results show that detectors operating on IPGs generalize across attack types and threat models, suggesting that provenance captures shared structural deviations that are attributed to adversarial inputs.

This paper makes the following contributions:
\begin{itemize}
\item \textbf{Inference Provenance as a Detection Signal.}
We formalize inference provenance as a graph-based representation of model execution and demonstrate its utility for adversarial example detection. Our results show that provenance captures cross-layer dependencies that are not accessible to layer-local methods.

\item \textbf{\sysname{} Extraction Framework.}
We develop a reproducible PyTorch-based extraction pipeline that instruments model execution via forward hooks and constructs heterogeneous IPGs. The framework operates at the level of tensor operations and module boundaries, requiring minimal model-specific adaptation.

\item \textbf{\sysname{}-IPG Dataset and Benchmark.}
We release a multi-domain dataset of IPGs spanning image classification and malware detection, including benign inputs and adversarial examples generated under both white-box and black-box threat models. The dataset includes standardized schemas, metadata, and extraction configurations to support reproducible research. The dataset is available at: \url{https://drive.google.com/file/d/1_uZE5c-dz5SnEd6S8V-8zDJed4ONdcQE/view?usp=sharing}

\item \textbf{Transferable Detection Across Attacks.}
Through cross-attack evaluations, we show that IPG-based detectors trained on a subset of attacks generalize to unseen ones, indicating that adversarial manipulation induces consistent structural deviations in inference behavior.
\end{itemize}

Reproducible code and datasets are available at:
{\small \url{https://github.com/um-dsp/NeuroTrace}}.
\section{Background}\label{sec:bground}

\subsection{Preliminaries}

\noindent\textbf{Supervised Learning.}
We consider a standard supervised learning setting with dataset $\mathcal{D} = \{(x_i, y_i)\}_{i=1}^{N}$, where $x_i \in \mathbb{R}^d$ and $y_i \in \{1, \dots, K\}$. A model $f_{\theta}: \mathbb{R}^d \rightarrow \mathbb{R}^K$, parameterized by $\theta$, is trained to minimize empirical risk:
\begin{equation}
    \mathcal{J}(\theta) = \frac{1}{N}\sum_{i=1}^{N}\mathcal{L}(f_{\theta}(x_i), y_i),
\end{equation}
where $\mathcal{L}$ is typically the cross-entropy loss. After training, $f_{\theta}$ is fixed and used during inference to produce predictions $\hat{y} = \arg\max_k [f_{\theta}(x)]_k$.

In this work, we do not modify the training process; instead, we analyze the \emph{inference-time execution} of $f_{\theta}$ for a given input $x$.

\medskip
\noindent\textbf{Graph Neural Networks (GNNs).}
A graph is defined as $G = (V, E)$, where $V$ is a set of nodes and $E$ is a set of edges. Each node $v \in V$ is associated with a feature vector $h_v$, and edges may also carry attributes. In \emph{heterogeneous graphs}, nodes and edges can belong to multiple types, defined by mappings $\phi: V \rightarrow \mathcal{A}$ and $\psi: E \rightarrow \mathcal{R}$.

GNNs learn representations by iteratively aggregating information from local neighborhoods:
\begin{equation}
    h_v^{(k+1)} = \sigma \left( \mathrm{AGG} \left( \{h_u^{(k)} : u \in \mathcal{N}(v)\} \right) \right),
\end{equation}
where $\mathcal{N}(v)$ denotes the neighbors of $v$.

In this work, GNNs are used as a downstream model to learn from graph-structured representations of inference. Our primary focus is on constructing informative graphs (IPGs), rather than proposing new GNN architectures.

\subsection{Inference Provenance Graphs (IPGs)}
\label{sec:ipgs}

We represent inference as a structured computational process using \emph{Inference Provenance Graphs (IPGs)}~\cite{hmida2025deepprovbehavioralcharacterizationrepair}.

\textbf{Definition.}
Let $f$ be a trained neural network and $x$ an input. The forward pass of $f(x)$ can be viewed as a sequence of tensor transformations defined over a directed acyclic graph (DAG). An IPG, denoted $G_x = (V_x, E_x)$, is an \emph{input-dependent subgraph} of this computation graph that captures the flow of information induced by $x$.

\textbf{Node Construction.}
Each node $v \in V_x$ corresponds to an intermediate activation generated during inference. Depending on the architecture, nodes may represent:
\begin{itemize}
    \item individual neurons (fully connected layers),
    \item channels (convolutional layers), or
    \item aggregated feature maps.
\end{itemize}
Each node is associated with a feature vector $h_v$ derived from its activation tensor (e.g., raw values, norms, or sparsity indicators). This design allows IPGs to accommodate heterogeneous architectures while maintaining a unified representation.

\textbf{Edge Construction.}
Edges encode computational dependencies between activations across layers. Formally, an edge $(u,v) \in E_x$ exists if the activation at node $u$ contributes to the computation of node $v$ during the forward pass. 

Edge attributes capture the nature of this dependency:
\begin{itemize}
    \item For parametric layers (e.g., linear, convolutional), edge attributes are derived from the corresponding learned weights.
    \item For non-parametric operations (e.g., pooling, normalization), edges encode structural connectivity without learned parameters.
\end{itemize}

\textbf{Activation-Based Subgraph Extraction.}
Although the model defines a full computation graph, not all nodes contribute equally to a given inference. We therefore construct IPGs as \emph{activation-induced subgraphs}. Specifically, a node $v$ is included in $V_x$ if its activation magnitude exceeds a threshold $\tau$:
\begin{equation}
    v \in V_x \quad \text{if} \quad \|h_v\| \geq \tau.
\end{equation}
This yields a compact representation that emphasizes the active computational pathways induced by input $x$.

\medskip
\noindent\textbf{Heterogeneous Representation.}
IPGs are naturally heterogeneous:
\begin{itemize}
    \item different node types correspond to different layer abstractions,
    \item edge types encode distinct computational relationships.
\end{itemize}
This heterogeneity allows the representation to preserve both structural and semantic distinctions across layers, which we later show to be important for detection performance.

\subsection{Activation-Induced Information Flow}
\label{sec:activation_dynamics}

The structure of an IPG is governed by activation dynamics during inference. Non-linearities such as ReLU induce sparsity by suppressing low-magnitude signals, effectively gating which nodes and edges participate in computation. As a result, each input induces a distinct execution pattern over the same model.

This perspective differs fundamentally from post-hoc attribution methods (e.g., LRP~\cite{binder2016layerwiserelevancepropagationneural}), which assign importance scores after inference. In contrast, IPGs are constructed directly from the forward pass and therefore reflect the \emph{actual runtime behavior} of the model.

\medskip
\noindent
We hypothesize that adversarial perturbations do not merely alter individual activations, but instead induce \emph{systemic changes in information flow} across the network. By explicitly modeling these patterns, IPGs provide a structured representation that can expose such deviations and serve as a signal for adversarial detection.
\section{\sysname{} Framework}
\label{sec:framework}

\begin{figure*}[t!]
    \centering   
    \includegraphics[scale=0.5]{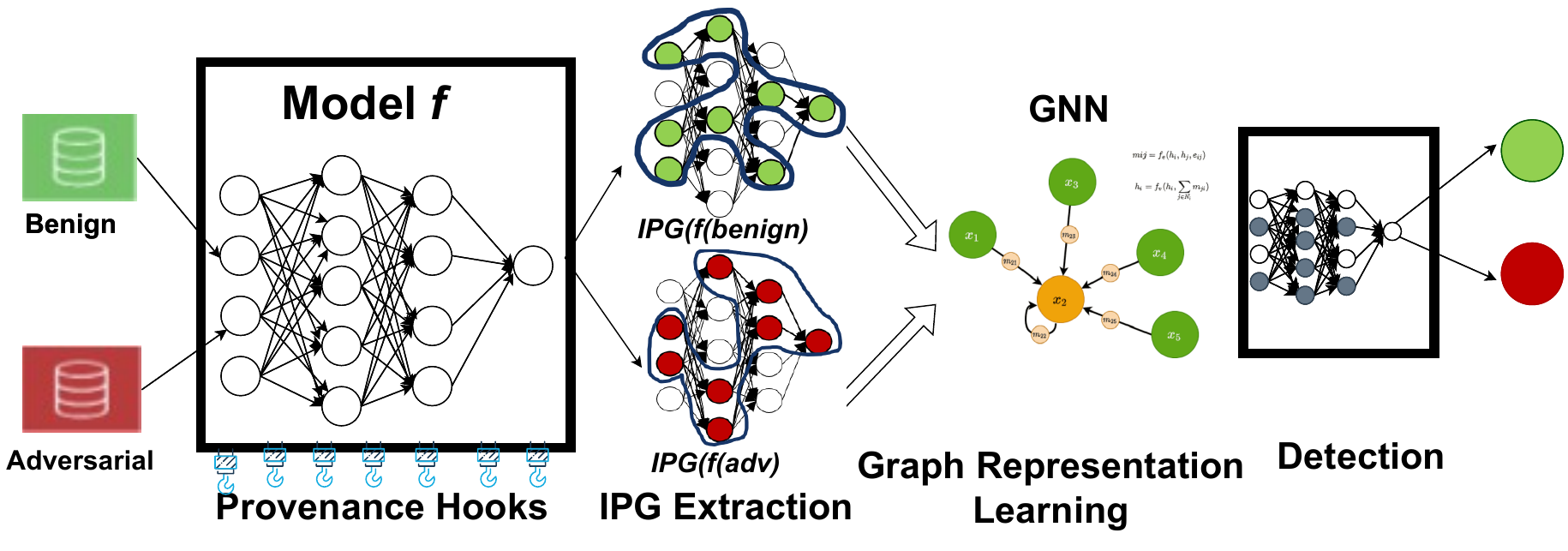}
    \caption{\sysname{} pipeline. Given a trained model and input, the framework extracts an input-dependent Inference Provenance Graph (IPG), represents it as a heterogeneous graph, and applies a graph classifier for downstream tasks such as adversarial detection.}
    \label{fig:tool}
    \vspace{-1em}
\end{figure*}

\sysname{} is a framework for \emph{constructing} and \emph{learning from} inference provenance representations. Given a trained model $f$ and an input $x$, the framework extracts an input-dependent graph $G_x$ that captures the model’s execution behavior, and applies a graph-based classifier for downstream analysis.

The pipeline consists of three components:
\begin{enumerate}
    \item \textbf{IPG Extraction:} Instrument the forward pass to construct an input-dependent graph representation.
    \item \textbf{Representation Learning:} Encode IPGs using a graph neural network.
    \item \textbf{Detection:} Predict whether an input exhibits adversarial characteristics.
\end{enumerate}

We emphasize that the primary contribution is the \emph{representation of inference provenance}; the GNN serves as a downstream model to evaluate its utility.

\subsection{IPG Extraction}
\label{sec:extraction}

\textbf{Instrumentation.}
Given a trained model $f$, \sysname{} attaches forward hooks to module boundaries (e.g., convolutional and linear layers) using a \texttt{ProvenanceEngine}. These hooks intercept activations during inference and record the dependencies required to construct the IPG.

This design operates at the level of tensor operations and does not assume a specific architecture, although the resulting graph granularity depends on how the model is modularized.

\medskip
\noindent\textbf{Node Construction.}
Each node $u$ represents an intermediate activation. To standardize representation across architectures, we use channel-level abstraction for convolutional layers and neuron-level abstraction for fully connected layers.

Each node stores:
\begin{itemize}
    \item \textbf{Activation features} $x_u$: summary statistics of the activation tensor (mean, $\ell_2$ norm, and sparsity ratio),
    \item \textbf{Activation mask} $m_u \in \{0,1\}$: indicating whether $\|x_u\| \geq \tau$,
    \item \textbf{Metadata}: layer index and layer type.
\end{itemize}

We use a threshold $\tau$ to filter low-magnitude activations; sensitivity to $\tau$ is evaluated in Appendix~X.

\medskip
\noindent\textbf{Edge Construction.}
Edges represent computational dependencies between layers. An edge $(u,v)$ is created if activation $u$ contributes to $v$ in the forward pass.

Edge attributes are defined as:
\begin{itemize}
    \item \textbf{Parametric layers:} edge weights derived from learned parameters,
    \item \textbf{Non-parametric layers:} unit-valued edges indicating structural connectivity.
\end{itemize}

For convolutional layers, we aggregate kernel weights at the channel level to maintain tractability.

\medskip
\noindent\textbf{Graph Representation.}
Each IPG is stored as a heterogeneous graph $G_x = (V_x, E_x)$ using PyTorch Geometric \texttt{HeteroData}. Node and edge types correspond to layer abstractions and inter-layer connections.

\medskip
\noindent\textbf{Dataset Construction.}
We construct datasets of labeled graphs:
\begin{equation}
\mathcal{D} = \{(G_i, y_i)\}, \quad y_i \in \{0,1\},
\end{equation}
where $y_i=1$ denotes adversarial inputs.

To prevent data leakage, splits are constructed at the \emph{input level}, ensuring no sample appears across train/validation/test sets.

\medskip
\noindent\textbf{Implementation Details.}
The extraction pipeline is implemented in PyTorch with:
\begin{itemize}
    \item deterministic seeds,
    \item configuration logging,
    \item Dockerized environments for reproducibility.
\end{itemize}

All extraction parameters (e.g., $\tau$, layer mappings) are explicitly specified in configuration files.

\subsection{Representation Learning and Detection}
\label{sec:learning}

\textbf{Feature Standardization.}
Since node feature dimensions vary across layers, we project all node features to a fixed dimension $d$ via zero-padding and linear projection.

\medskip
\noindent\textbf{GNN Architecture.}
We use a 3-layer GraphSAGE model with:
\begin{itemize}
    \item hidden dimension: 128,
    \item mean aggregation,
    \item ReLU activations,
    \item global mean pooling for graph-level representation.
\end{itemize}

We also evaluate GCN and GAT variants in Appendix~X.

\medskip
\noindent\textbf{Training.}
The model is trained with binary cross-entropy loss using Adam (learning rate $10^{-3}$). Mini-batches consist of $B$ graphs.

\medskip
\noindent\textbf{Detection.}
Given a graph $G_x$, the trained model outputs $p(y=1|G_x)$, representing the likelihood of adversarial manipulation.

\subsection{Runtime and Storage Analysis}
\label{sec:complexity}

\textbf{Extraction Cost.}
Extraction scales linearly with graph size:
\begin{equation}
    T = \Theta(|V_x| + |E_x|).
\end{equation}

On a single NVIDIA V100 GPU:
\begin{itemize}
    \item ResNet-20 (CIFAR-10): $\sim$15--25 ms per sample (batched),
    \item EMBER-DNN: $\sim$5--10 ms per sample.
\end{itemize}

\medskip
\noindent\textbf{Storage Cost.}
Each IPG requires:
\begin{itemize}
    \item $\sim$0.5--2 MB (CIFAR-10),
    \item $\sim$0.2--0.8 MB (malware models),
\end{itemize}
depending on graph sparsity and feature dimensions.

\medskip
\noindent\textbf{Scalability Considerations.}
To scale to larger models, we support:
\begin{itemize}
    \item channel-level aggregation,
    \item activation thresholding,
    \item optional layer subsampling.
\end{itemize}

These reduce graph size while preserving structural information. Extending to large architectures (e.g., ViTs) is discussed in Section~X.

\section{IPG Dataset Organization}
\label{sec:dataset}

\textbf{Overview.}
We release a dataset of IPGs spanning vision and malware domains. Each graph corresponds to a single inference event.

\medskip
\noindent\textbf{Storage Format.}
Graphs are stored as PyTorch Geometric \texttt{HeteroData} objects:
\begin{description}
    \item[\texttt{graphs/}] Serialized IPGs, one per input.
\end{description}

\medskip
\noindent\textbf{Metadata.}
Each graph is associated with:
\begin{itemize}
    \item model identifier,
    \item input label,
    \item attack type (if applicable),
    \item split assignment,
    \item extraction configuration.
\end{itemize}

\medskip
\noindent\textbf{Dataset Splits.}
We provide train/validation/test splits with strict separation by input sample. Additionally, we include cross-attack splits for evaluating transferability.

\medskip
\noindent\textbf{Statistics.}
For each dataset, we report:
\begin{itemize}
    \item average node/edge counts,
    \item activation sparsity,
    \item graph size distribution.
\end{itemize}

These statistics are included in a manifest file for reproducibility.

\medskip
\noindent
The dataset is designed to support both provenance-based detection and broader analysis of inference-time behavior.
\section{Validation Experiments}
\label{sec:experiments}

We evaluate \sysname{} on \emph{adversarial input detection}: given a trained classifier $f$ and an input $x$, the goal is to determine whether the corresponding inference provenance graph $G_x$ was induced by a benign or adversarial example. Our evaluation is designed to answer four questions:

\begin{enumerate}
    \item \textbf{Intra-attack performance:} Can IPG-based detectors distinguish benign and adversarial inputs when trained and tested on the same attack family?
    \item \textbf{Cross-attack transfer:} Do provenance-based detectors generalize to unseen attack families?
    \item \textbf{Cross-threat transfer:} Do representations learned from white-box attacks transfer to black-box attacks, and vice versa?
    \item \textbf{Practicality:} What are the runtime and storage costs of extracting and storing IPGs?
\end{enumerate}

\begin{figure}[t!]
    \centering   
    \includegraphics[scale=0.45]{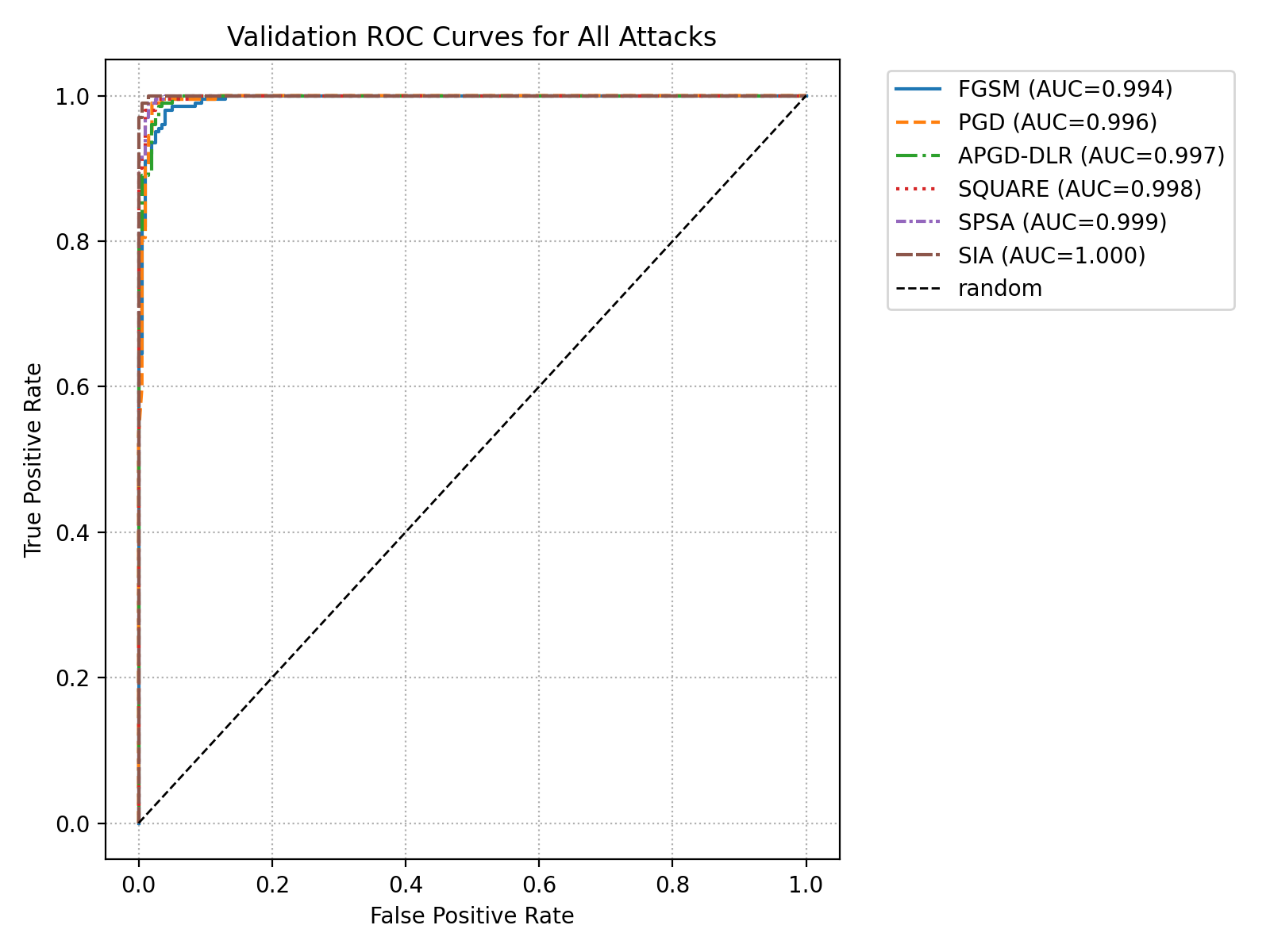}
    \caption{\textbf{ROC curves under intra-attack evaluation.} The detector is trained and tested on the same attack family. Across attacks, the curves remain close to the upper-left corner, indicating strong separability between benign and adversarial IPGs.}
    \label{fig:all_attacks_val_roc}
\end{figure}

\begin{figure}[t!]
    \centering   
    \includegraphics[scale=0.45]{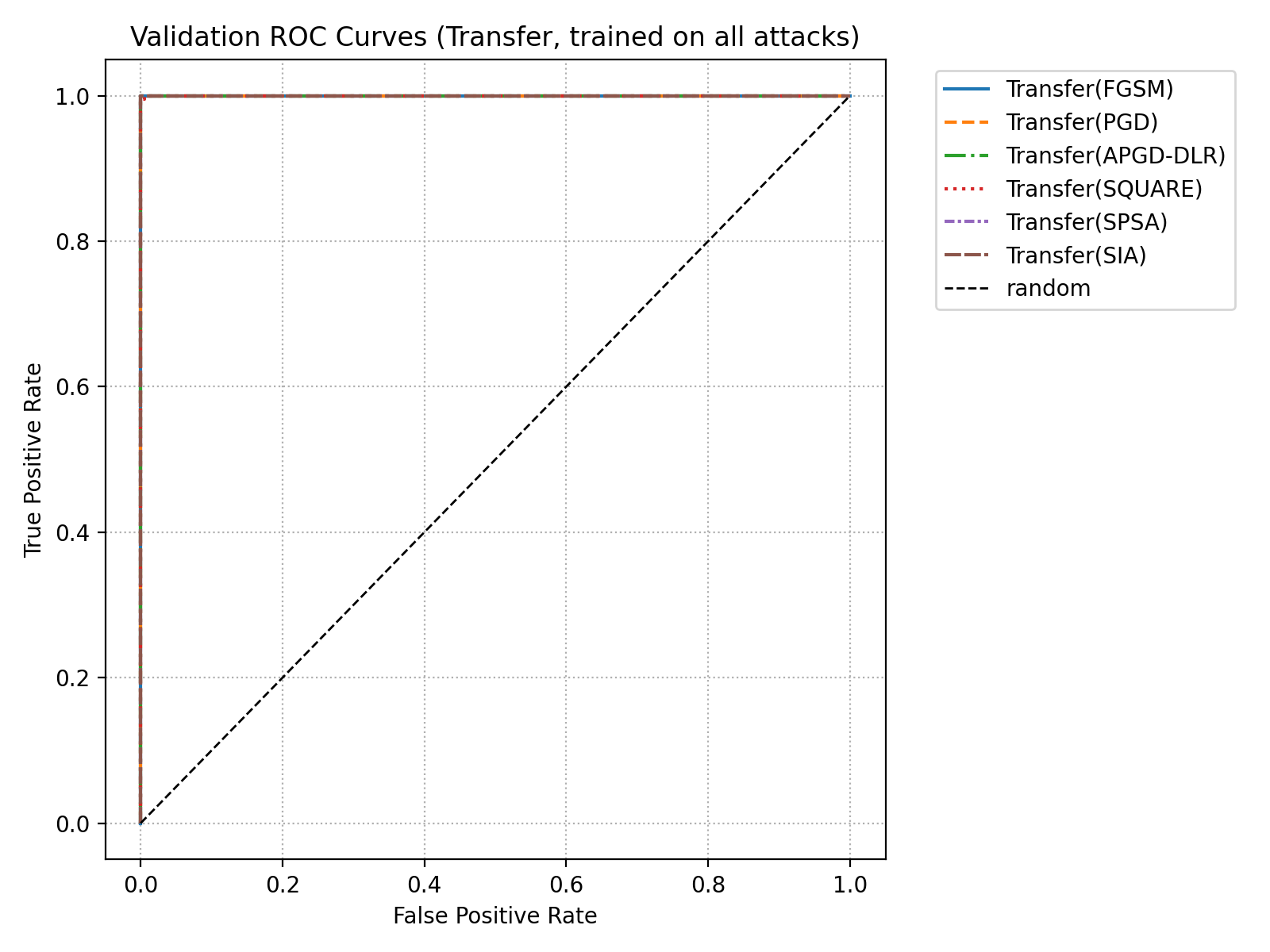}
    \caption{\textbf{ROC curves under multi-attack training.} A single detector is trained on a mixture of attacks and evaluated separately on each attack type. The consistently strong ROC profiles suggest that inference provenance captures structural signals that are shared across attack families.}
    \label{fig:transfer_all_attacks_val_roc}
\end{figure}

Figures~\ref{fig:all_attacks_val_roc} and~\ref{fig:transfer_all_attacks_val_roc} summarize the resulting ROC behavior in the intra-attack and multi-attack settings, respectively.

\subsection{Experimental Setup}
\label{sec:exp_setup}

\noindent\textbf{Target Models and Domains.}
We evaluate \sysname{} on three trained classifiers spanning two domains:
\begin{itemize}
    \item \textbf{Vision:} a ResNet-20 classifier trained on CIFAR-10~\cite{CIFAR10} with test accuracy of approximately $95\%$.
    \item \textbf{Malware:} two fully connected malware detectors, \texttt{Cuckoo-DNN} trained on Cuckoo-Traces~\cite{cuckoo-data} and \texttt{EMBER-DNN} trained on EMBER~\cite{EMBER2018}, with accuracies of $95.48\%$ and $94.56\%$, respectively.
\end{itemize}
These models were selected to cover both convolutional and dense architectures, thereby testing whether the extraction pipeline and downstream provenance representation remain useful across distinct modalities and network structures.

\medskip
\noindent\textbf{Adversarial Attacks.}
For CIFAR-10, we consider six attacks: three white-box attacks---FGSM~\cite{FGSM}, PGD~\cite{PGD}, and APGD-DLR from AutoAttack~\cite{APGD-DLR}---and three black-box attacks---SPSA~\cite{uesato2018adversarial}, Square~\cite{andriushchenko2020square}, and SIA/SIT~\cite{wang2023structure}. To avoid naming ambiguity, we use a single notation consistently throughout the paper; specifically, the attack listed as \emph{SIA} in Tables~\ref{tab:attack_eval}--\ref{tab:attack_eval_w2bb2w} corresponds to the same structure-aware black-box attack referred to elsewhere as \emph{SIT}. In the final version, we standardize all references to one name.

For CIFAR-10, all perturbations are generated under an $\ell_{\infty}$ budget of $\epsilon \leq 0.3$ on inputs normalized to $[0,1]$. This is a relatively strong perturbation regime and is intended to stress-test whether provenance signatures remain separable under substantial input manipulation. We note that this choice should be interpreted under the stated normalization convention.

For the malware models, we evaluate two feature-space attacks already used in our benchmark construction:
\begin{itemize}
    \item \textbf{Emb-att} against the EMBER classifier, and
    \item \textbf{Bit-Flip} against the Cuckoo-based classifier.
\end{itemize}
These attacks perturb tabular or program-behavior features rather than pixels, allowing us to assess whether provenance-based detection extends beyond image classification.

\begin{figure}[t]
    \centering   
    \includegraphics[scale=0.33]{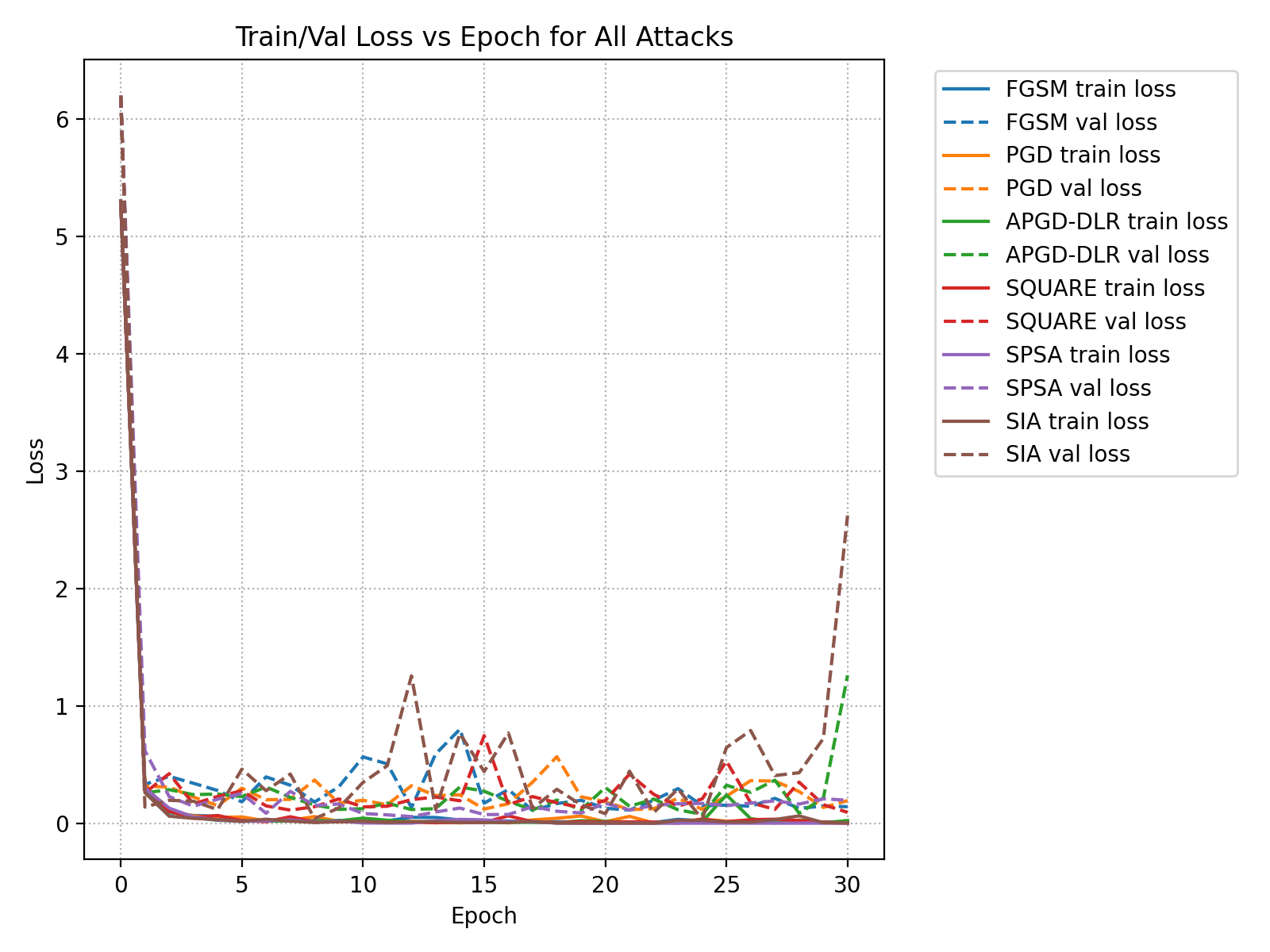}
    \caption{\textbf{Training and validation loss} for the multi-attack detector. The curves indicate stable optimization without evidence of divergence.}
    \label{fig:train-val_loss}
\end{figure}

\begin{figure}[t]
    \centering   
    \includegraphics[scale=0.33]{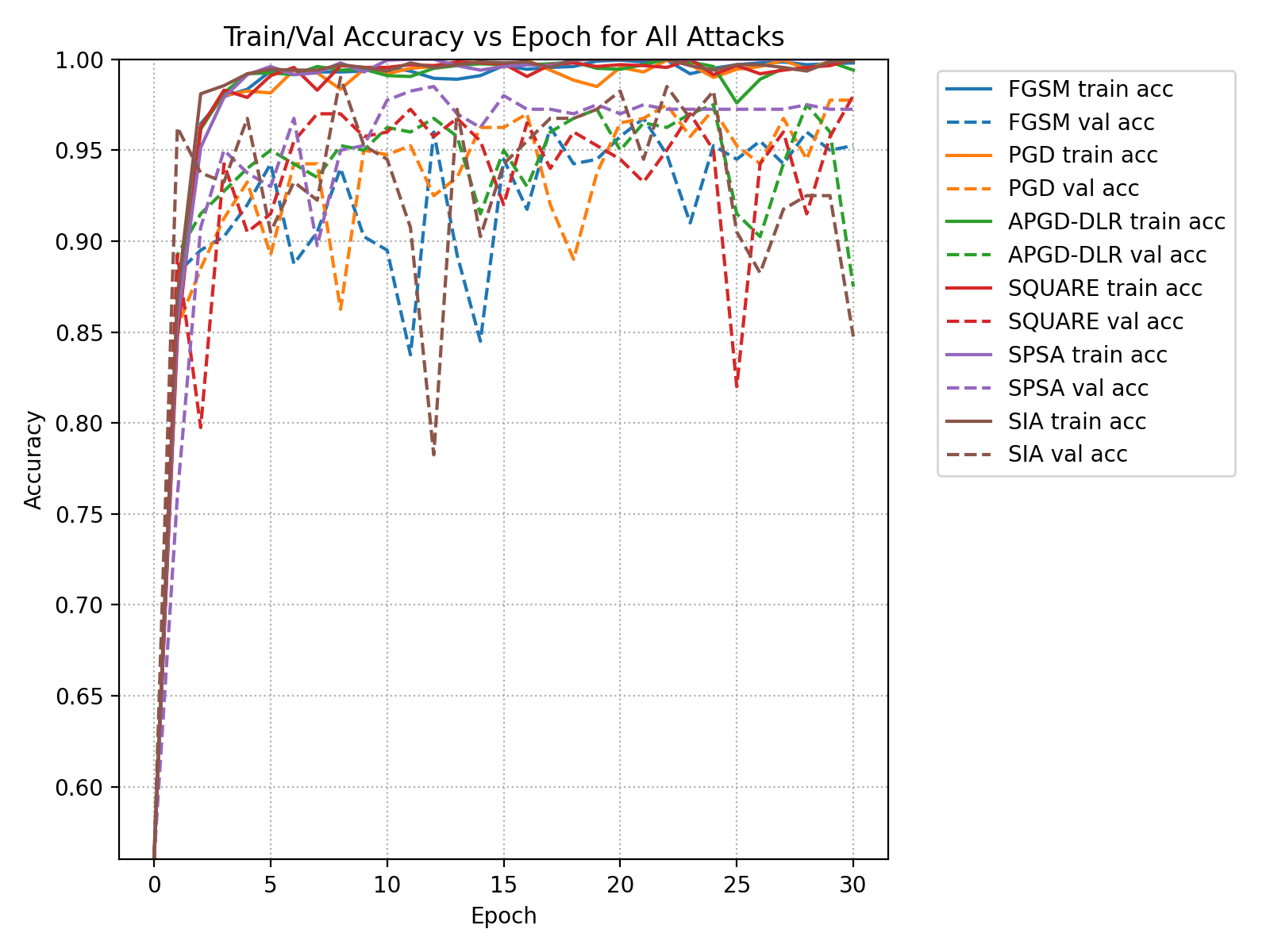}
    \caption{\textbf{Training and validation accuracy} for the multi-attack detector. Accuracy improves rapidly and remains stable, indicating that the learned provenance representation is easy to separate under supervised training.}
    \label{fig:train-val-acc}
\end{figure}

\medskip
\noindent\textbf{Detector Architecture and Training.}
Unless otherwise noted, the detector is the heterogeneous GNN described in Section~\ref{sec:learning}. We train with binary cross-entropy loss using Adam, and evaluate on held-out graphs only. Training dynamics for the multi-attack detector are shown in Figures~\ref{fig:train-val_loss} and~\ref{fig:train-val-acc}; both indicate stable optimization and fast convergence.

\subsection{Evaluation Protocol}
\label{sec:exp_protocol}

We structure the evaluation into three complementary protocols.

\medskip
\noindent\textbf{(1) Intra-Attack Evaluation.}
For each attack family, we train the detector on benign and adversarial IPGs from that attack alone and test on a held-out split from the same attack. For CIFAR-10, we use an $80/20$ split with 800 benign and 800 adversarial training graphs, and 200 benign and 200 adversarial test graphs per attack.

\medskip
\noindent\textbf{(2) Multi-Attack Training and Per-Attack Testing.}
We train a single detector on a mixture of attack families and evaluate it separately on each attack. This setting does \emph{not} test strict unseen-attack generalization; rather, it measures whether a single provenance-based detector can jointly model multiple attack distributions without attack-specific retraining. Because the detector sees examples from all attacks during training, we avoid calling this setting ``attack-agnostic'' and instead interpret it as \emph{multi-attack robustness}.

\medskip
\noindent\textbf{(3) Cross-Threat Transfer.}
To test transferability more directly, we train on only white-box attacks and evaluate on black-box attacks, and then reverse the direction. This protocol probes whether IPG-based detectors learn structural signals that transcend the attack-generation mechanism.

\medskip
\noindent\textbf{(4) Baseline Comparison.}
We compare \sysname{} against CIGA~\cite{CIGA}, which is the closest graph-based baseline available in our experimental setting. We report side-by-side detection performance on CIFAR-10.

\subsection{Metrics}
\label{sec:exp_metrics}

We report standard binary classification metrics: Accuracy, Precision, Recall, and F1-score. To assess threshold-independent ranking quality, we also report ROC-AUC and PR-AUC. Because adversarial detection is a security-sensitive setting in which low false-positive operation is important, we additionally report:
\begin{itemize}
    \item \textbf{TPR@1\%FPR}: true positive rate at $1\%$ false positive rate,
    \item \textbf{FPR@95\%TPR}: false positive rate at $95\%$ true positive rate.
\end{itemize}

For a few malware settings with near-perfect score separation, the empirical ROC curve can become degenerate at extreme operating points, which makes TPR@1\%FPR unstable or uninformative even when overall detection is perfect. We therefore interpret such entries jointly with ROC-AUC, PR-AUC, Accuracy, and F1 rather than in isolation.

\subsection{Results}
\label{sec:exp_results}

\subsubsection{Intra-Attack Detection Performance}

Table~\ref{tab:attack_eval} reports the results when the detector is trained and tested on the same attack family. On CIFAR-10, \sysname{} achieves uniformly strong performance across both white-box and black-box attacks, with all Accuracy, F1, ROC-AUC, and PR-AUC values above $96\%$ and most AUC values approaching $1.0$. These results indicate that IPG-level features carry strong discriminative signal for separating benign and adversarial inference events.

\begin{table*}[t!]
\centering
\caption{IPG-based detection performance under intra-attack evaluation. Each detector is trained and tested on the same attack family.}
\label{tab:attack_eval}
\scalebox{0.85}{
\begin{tabular}{lcccccccc}
\toprule
\textbf{Attack} &
\textbf{Accuracy} &
\textbf{Precision} &
\textbf{Recall} &
\textbf{F1} &
\textbf{ROC-AUC} &
\textbf{PR-AUC} &
\textbf{TPR@1\%FPR} &
\textbf{FPR@95\%TPR} \\
\midrule
FGSM     & 96.75 & 96.06 & 97.50 & 96.77 & 99.45 & 99.41 & 91.0 & 2.5 \\
PGD      & 97.75 & 96.14 & 99.50 & 97.79 & 99.56 & 99.49 & 90.5 & 2.0 \\
APGD-DLR & 97.50 & 96.12 & 99.00 & 97.54 & 99.68 & 99.67 & 89.0 & 2.0 \\
Square   & 98.00 & 96.60 & 99.50 & 98.03 & 99.85 & 99.85 & 98.0 & 1.0 \\
SPSA     & 98.50 & 98.02 & 99.00 & 98.51 & 99.88 & 99.88 & 97.0 & 1.0 \\
SIA      & 99.00 & 98.04 & 100.00 & 99.01 & 99.98 & 99.97 & 99.0 & 0.0 \\
Emb-att  & 100.00 & 100.00 & 100.00 & 100.00 & 100.00 & 100.00 & 0.0 & 0.0 \\
Bit-Flip & 100.00 & 100.00 & 100.00 & 100.00 & 100.00 & 100.00 & 0.0 & 0.0 \\
\bottomrule
\end{tabular}}
\end{table*}

The malware-domain results are also strong: both \texttt{Emb-att} and \texttt{Bit-Flip} achieve perfect separation under the current benchmark. This is important because it demonstrates that provenance-based detection is not confined to image models. At the same time, these results should be interpreted cautiously: the current malware evaluation covers fewer attack families and therefore provides evidence of \emph{domain applicability}, but not yet the same breadth of robustness analysis as the CIFAR-10 experiments.

At a mechanistic level, the strong intra-attack performance suggests that adversarial perturbations induce persistent deviations in internal execution structure, not merely changes in the final prediction. Because the detector operates on graph-structured inference traces, it is able to exploit distortions that are distributed across layers.

\subsubsection{Multi-Attack Training}

Table~\ref{tab:attack_eval_all} evaluates a single detector trained on a mixture of all CIFAR-10 attacks and tested on each attack separately. Performance is nearly saturated across all metrics. These results show that one detector can jointly capture multiple known attack distributions without attack-specific retraining.

\begin{table*}[t!]
\centering
\caption{IPG-based detection performance when training on a mixture of all CIFAR-10 attacks and evaluating each attack separately. Because training includes all attack families, these results measure multi-attack robustness rather than unseen-attack generalization.}
\label{tab:attack_eval_all}
\scalebox{0.85}{
\begin{tabular}{lcccccccc}
\toprule
\textbf{Attack} &
\textbf{Accuracy} &
\textbf{Precision} &
\textbf{Recall} &
\textbf{F1} &
\textbf{ROC-AUC} &
\textbf{PR-AUC} &
\textbf{TPR@1\%FPR} &
\textbf{FPR@95\%TPR} \\
\midrule
Transfer(FGSM)          & 99.75 & 100.00 & 99.50 & 99.75 & 100.00 & 100.00 & 100.0 & 0.0 \\
Transfer(PGD)           & 99.75 & 100.00 & 99.50 & 99.75 & 100.00 & 100.00 & 100.0 & 0.0 \\
Transfer(Auto-Attack)   & 99.75 & 100.00 & 99.50 & 99.75 & 100.00 & 100.00 & 100.0 & 0.0 \\
Transfer(Square)        & 99.75 & 100.00 & 99.50 & 99.75 & 100.00 & 100.00 & 100.0 & 0.0 \\
Transfer(SPSA)          & 100.00 & 100.00 & 100.00 & 100.00 & 100.00 & 100.00 & 100.0 & 0.0 \\
Transfer(SIT)           & 100.00 & 100.00 & 100.00 & 100.00 & 100.00 & 100.00 & 100.0 & 0.0 \\
\bottomrule
\end{tabular}}
\end{table*}

Importantly, these numbers do \emph{not} by themselves establish that the detector is attack-agnostic in the strict sense, since the training set contains samples from all evaluated attacks. Instead, they support a more precise conclusion: inference provenance provides a sufficiently expressive representation for building a \emph{single, jointly trained detector} that handles multiple known attack families.

\subsubsection{Cross-Threat Transfer}

Table~\ref{tab:attack_eval_w2bb2w} presents the stricter transfer setting: training on white-box attacks and testing on black-box attacks, and vice versa. Here the detector still performs strongly, with Accuracy generally above $96\%$ and near-perfect AUC values. This is the strongest evidence in the paper that provenance captures structural regularities that transfer across threat models.

\begin{table*}[t!]
\centering
\caption{Cross-threat transfer performance. The detector is trained on white-box attacks and tested on black-box attacks, and vice versa.}
\label{tab:attack_eval_w2bb2w}
\scalebox{0.85}{
\begin{tabular}{lcccccccc}
\toprule
\textbf{Attack} &
\textbf{Accuracy} &
\textbf{Precision} &
\textbf{Recall} &
\textbf{F1} &
\textbf{ROC-AUC} &
\textbf{PR-AUC} &
\textbf{TPR@1\%FPR} &
\textbf{FPR@95\%TPR} \\
\midrule
Black-Box$\rightarrow$White-Box(FGSM)        & 100.00 & 100.00 & 100.00 & 100.00 & 100.00 & 100.00 & 100.0 & 0.0 \\
Black-Box$\rightarrow$White-Box(PGD)         & 96.67  & 100.00 & 93.33  & 96.55  & 99.89 & 99.89 & 96.67 & 0.0 \\
Black-Box$\rightarrow$White-Box(Auto-Attack) & 100.00 & 100.00 & 100.00 & 100.00 & 100.00 & 100.00 & 100.0 & 0.0 \\
White-Box$\rightarrow$Black-Box(Square)      & 98.33 & 96.77 & 100.00 & 98.36 & 99.89 & 99.89 & 96.67 & 0.0 \\
White-Box$\rightarrow$Black-Box(SPSA)        & 98.33 & 96.77 & 100.00 & 98.36 & 99.78 & 99.78 & 93.33 & 3.33 \\
White-Box$\rightarrow$Black-Box(SIT)         & 98.33 & 96.77 & 100.00 & 98.36 & 100.00 & 100.00 & 100.00 & 0.0 \\
\bottomrule
\end{tabular}}
\end{table*}

This transferability is consistent with the intuition behind IPGs. White-box and black-box attacks differ substantially in how they are generated, yet both ultimately aim to steer the model toward incorrect behavior. If adversarial manipulation systematically alters internal execution, then provenance-based detectors should remain effective even when the perturbation mechanism changes. Our results support this view.

\subsubsection{Comparison with Prior Work}

Table~\ref{tab:sota_cifar10} compares \sysname{} with CIGA~\cite{CIGA}, the strongest graph-based baseline available in our setup. \sysname{} outperforms CIGA across all reported attacks, with especially large gains on PGD and APGD.

\begin{table}[t!]
\centering
\caption{Comparison with CIGA on CIFAR-10. Detection rate is reported in percent.}
\label{tab:sota_cifar10}
\scalebox{0.8}{
\begin{tabular}{lcccccc}
\toprule
\textbf{Method} & \textbf{FGSM} & \textbf{PGD} & \textbf{APGD} & \textbf{SPSA} & \textbf{SIT} & \textbf{Square} \\
\midrule
CIGA~\cite{CIGA}      & 99.49 & 90.09 & 90.01 & 93.56 & 95.62 & 94.16 \\
\sysname{} (Ours)     & \textbf{99.75} & \textbf{99.75} & \textbf{99.75} & \textbf{98.89} & \textbf{98.33} & \textbf{98.61} \\
\bottomrule
\end{tabular}}
\end{table}

The comparison should be interpreted in context. CIGA is a graph-based detector and therefore a natural baseline for evaluating whether richer provenance structure improves detection. A broader comparison against non-graph-based detectors would further strengthen the empirical picture and remains an important next step.

\subsection{Runtime and Storage Overhead}
\label{sec:runtime_storage_eval}

Table~\ref{tab:ipg_runtime_storage_avg} reports the empirical overhead of IPG extraction and storage. We report average end-to-end extraction time per graph, average graph size, and average serialized size. These results make clear that provenance capture is not free: mid-sized vision models can require tens of seconds per graph under the current extraction implementation, whereas smaller malware models are substantially cheaper.

\begin{table}[t]
\centering
\caption{\textbf{Empirical runtime and storage cost of IPG extraction.} $T_{\mathrm{overhead}}$ denotes average end-to-end wall-clock extraction time per graph. $N$ and $E$ are average node and edge counts. $S_{\mathrm{IPG}}$ is average serialized graph size.}
\label{tab:ipg_runtime_storage_avg}
\begin{adjustbox}{max width=\linewidth}
\begin{tabular}{llcccccc}
\toprule
\textbf{Type} & \textbf{Attack} &
\textbf{$T_{\mathrm{overhead}}$ (s)} &
\textbf{$N$} &
\textbf{$E$} &
\textbf{$d_v$} &
\textbf{$d_e$} &
\textbf{$S_{\mathrm{IPG}}$ (MB/graph)} \\
\midrule
Resnet20     & FGSM        & 21.25 & 2330 & 108909 & 32 & 1 & 5.30 \\
Resnet20     & PGD         & 15.99 & 2330 & 108755 & 32 & 1 & 5.30 \\
Resnet20     & SQUARE      & 16.71 & 2330 & 108737 & 32 & 1 & 5.30 \\
\midrule
Ember-Model  & Emb-att MLP & 24.64 & 2702 & 126307 & 32 & 1 & 6.15 \\
Cuckoo-Model & BitFlip     & 3.93  & 431  & 20159  & 32 & 1 & 0.98 \\
\bottomrule
\end{tabular}
\end{adjustbox}
\end{table}

Two conclusions follow. First, the extraction cost scales with graph size, as expected from the complexity analysis in Section~\ref{sec:complexity}. Second, although the current implementation is practical for offline analysis, benchmarking, and high-assurance settings, further compression and selective-capture mechanisms will be needed for low-latency deployment on larger architectures. These measurements therefore support both the practicality and the current limitations of provenance extraction.

\subsection{Discussion}
\label{sec:discussion}

\noindent\textbf{Inference provenance as a detection signal.}
Across vision and malware settings, the results consistently indicate that inference provenance contains rich information for identifying adversarial behavior. Unlike detectors that rely only on final-layer confidence or localized activation summaries, \sysname{} operates on the full execution trace induced by an input. This allows it to capture changes in inter-layer dependencies and activation structure that persist across attack families.

\medskip
\noindent\textbf{Why provenance transfers.}
The strongest evidence for the usefulness of provenance comes from the cross-threat transfer results. Training on white-box attacks and testing on black-box attacks remains effective, and vice versa. This suggests that adversarial examples generated by different mechanisms still induce a shared class of structural deviations in execution behavior. In that sense, provenance appears to expose a more global notion of inference abnormality than purely attack-specific artifacts.

\medskip
\noindent\textbf{What the current results do \emph{not} show.}
Several boundaries are important. First, the multi-attack results in Table~\ref{tab:attack_eval_all} should not be interpreted as evidence against a fully adaptive adversary, since the detector is trained on all attack families evaluated there. Second, we do not yet evaluate adaptive attacks that explicitly optimize to both fool the base model and evade the provenance detector. Such attacks are a crucial next step for assessing robustness under stronger threat models. Third, while the malware results are promising, they currently cover fewer attacks and therefore provide narrower evidence than the CIFAR-10 study.

\medskip
\noindent\textbf{Mechanistic interpretation.}
A plausible explanation for the observed performance is that adversarial perturbations induce what may be viewed as \emph{systemic tainting} of inference: although the perturbation may be small in input space, its effect propagates through many layers, changing activation patterns and connectivity structure in ways that are detectable at the graph level. This interpretation is consistent with recent observations that adversarial manipulation produces distributed internal shifts even when the perturbation is visually or semantically subtle~\cite{hmida2025deepprovbehavioralcharacterizationrepair}.

\medskip
\noindent\textbf{Practical implications.}
The overhead analysis indicates that provenance capture is currently best suited to offline auditing, forensic analysis, incident response, and other high-assurance settings in which additional latency is acceptable. In such contexts, IPGs provide more than a binary detection output: they also constitute a persistent artifact that can be archived, compared across events, and analyzed post hoc for debugging and forensics.

\medskip
\noindent\textbf{Limitations and future directions.}
This work focuses on supervised graph-level detection of adversarial inputs. Several directions remain open:
\begin{itemize}
    \item evaluation against adaptive attacks that explicitly target the provenance detector,
    \item broader comparisons against non-graph-based adversarial detectors,
    \item ablations isolating the role of node features, edge attributes, and graph heterogeneity,
    \item extension to larger architectures such as ViTs and other modern networks,
    \item graph compression and selective provenance capture for lower-latency deployment.
\end{itemize}
Overall, the present results establish that inference provenance is a useful and transferable signal for adversarial detection, while also clarifying the conditions under which that conclusion should be interpreted.
\section{Related Work}\label{sec:related}
We position \sysname{} with respect to previous work
on detecting adversarial samples. Existing methods are broadly categorized into supervised and unsupervised approaches~\cite{ma2018lid,kherchouche2020nss,sotgiu2019dnr,ma2019nic}. 
Supervised detectors such as LID~\cite{ma2018lid} and NSS~\cite{kherchouche2020nss} design discriminatory features to train classifiers that separate benign from adversarial samples~\cite{ma2018lid,kherchouche2020nss}. 
However, these methods can be sensitive to attack configurations and may generalize poorly to unseen or transferred attacks. 
Unsupervised approaches such as DNR~\cite{sotgiu2019dnr} and NIC~\cite{ma2019nic} mainly perform inconsistency or invariant checks using hidden-layer behaviors to identify adversarial inputs~\cite{sotgiu2019dnr,ma2019nic}. 
Overall, most existing works rely on data or activation features and emphasize per-layer effects, while inter-layer correlations are under-explored, and NIC's derived-model approximations may introduce redundancy that can hinder detection~\cite{ma2019nic}.

Recently, Zhang et al.~\cite{CIGA} propose Critical Inference Graphs (CIGs) to extract subgraphs that are critical to specific predictions, using LRP to identify important nodes for benign inputs and training one-class classifiers for anomaly detection. 
In contrast, \sysname{} constructs IPGs to capture broader inference-time behavior and dependencies beyond only LRP-selected nodes. 
By modeling richer cross-layer and runtime-level relations, our IPG-based design reduces reliance on a single attribution filter and enables detection signals that extend beyond pure classification-critical substructures. 
Thus, \sysname{} complements CIGs by offering a more holistic graph abstraction of DNN inference that can expose patterns missed when ignoring nodes deemed irrelevant by LRP.

\section{Conclusion}\label{sec:concl}

This work introduced \sysname{}, a unified framework and open dataset for analyzing inference provenance through Inference Provenance Graphs (IPGs). By capturing both activation dynamics and dataflow dependencies across layers, IPGs provide a structured and semantically rich representation of model behavior that extends beyond conventional attribution or layer-local activation analysis. \sysname{} contributes (i) a reproducible extraction engine, (ii) a standardized heterogeneous graph representation of inference, and (iii) a benchmark suite spanning adversarial examples in both vision and malware domains.

Our evaluation shows that IPG-based detectors achieve consistently strong performance across a range of attack families and settings, including intra-attack evaluation, multi-attack training, and cross-threat transfer between white-box and black-box attacks. These results suggest that adversarial perturbations induce detectable shifts in inference-time computational structure, and that such shifts can be captured through provenance representations. At the same time, the results highlight important distinctions between settings in which the detector is trained on known attacks and those that require generalization to unseen ones, reinforcing the need to interpret ``attack-agnostic'' behavior carefully.

Beyond detection performance, \sysname{} provides a reproducible and extensible substrate for studying inference-time behavior in machine learning systems. By releasing the extraction pipeline, dataset, graph schema, and evaluation protocol, we aim to support systematic investigation of provenance-based methods across tasks and domains.

Several limitations and opportunities for future work remain. In particular, we do not evaluate adaptive adversaries that explicitly optimize to evade provenance-based detectors, and our current experiments are limited to mid-sized models. Extending provenance extraction and analysis to larger architectures (e.g., ViTs and LLMs), conducting detailed ablation studies, and improving the efficiency of graph construction are important next steps. More broadly, integrating inference provenance with complementary signals such as uncertainty estimation or runtime monitoring may further enhance robustness.

Overall, our findings indicate that inference provenance is a practical and informative lens for analyzing model behavior under adversarial manipulation, and that IPGs provide a useful foundation for building more transparent and auditable machine learning systems.

\bibliographystyle{ACM-Reference-Format}
\bibliography{main}

\end{document}